\documentclass[11pt]{amsart}
\usepackage{amscd}
\usepackage{amsmath}
\usepackage{graphicx}
\usepackage{amsfonts}
\usepackage{amssymb}
\begin{document}

\title[Concurrence for infinite-dimensional systems]
{Concurrence for infinite-dimensional quantum systems}

\author{Yu Guo}
\address[Y. Guo]
{Department of Mathematics, Shanxi Datong University, Datong 037009,
China; \\
Institute of Optoelectroincs Engineering, Department of Physics and
Optoelectroincs, Taiyuan University of Technology, Taiyuan 030024,
China} \email{guoyu3@yahoo.com.cn}

\author{Jinchuan Hou}
\address[J. Hou]{Department of
Mathematics\\
Taiyuan University of Technology\\
 Taiyuan 030024,
  P. R. China}
\email{jinchuanhou@yahoo.com.cn, houjinchuan@tyut.edu.cn}
\author{Yuncai Wang}
\address[Y. Wang]
{Institute of Optoelectroincs Engineering, Department of Physics and
Optoelectroincs, Taiyuan University of Technology, Taiyuan 030024,
China} \email{2006wang.yinzhu@163.com}



\maketitle

\begin{abstract}
Concurrence is an important entanglement measure for states in
finite-dimensional quantum systems that was explored intensively in
the last decade. In this paper, we extend the concept of concurrence
to infinite-dimensional bipartite   systems and show that it is
continuous and does not increase under local operation and classical
communication (LOCC). Moreover, based on the partial Hermitian
conjugate (PHC) criterion proposed in [Chin. Phys. Lett.
\textbf{26}, 060305(2009); Chin. Sci. Bull. \textbf{56}(9),
840--846(2011)], we introduce a concept of the PHC measure   and
show that it coincides with the concurrence, which provides another
perspective on the concurrence.
\end{abstract}


\section{Introduction}
 Entanglement, being viewed as one of the key features
of quantum world that has no classical counterpart, is perhaps
\emph{the most challenging subject of modern quantum theory}. There
are two distinct directions for characterizing entanglement. One is
to find proper criteria of detecting entanglement, and the other is
to find a ``good'' entanglement measure, namely, to define the best
measure quantifying an amount of entanglement of a given state.
Among a number of entanglement measures, \emph{concurrence} is a
subject of intense research interest
\cite{Hill,Wootters,Wootters2,Gao,Chen,Fan,Rungta,Albeverio,Zhang,Chattopadhyay,Chattopadhyay2,Chen2,Berrada,Huang,Jafarpour,Salimi,Augusiak,Li},
which has been shown to play a key role in analyzing the ultrabright
source of entangled photon pairs \cite{Dousse}, describing quantum
phase transitions in various interacting quantum many-body systems
\cite{Osterloh}, affecting macroscopic properties of solids
significantly \cite{Ghosh}, exploring dynamics of entanglement for
noisy qubits that make diploe-diploe interaction \cite{Altintas} and
revealing distinct scaling behavior for different types of
multipartite entanglement \cite{Carvalho}, etc.

Concurrence is originally derived from the entanglement of formation
(EOF) which is used to compute the amount of entanglement for pure
states in two-qubit systems \cite{Hill}. Because of the EOF is a
monotonically increasing function of the concurrence, thus the
concurrence itself can also be regarded as an entanglement measure.
Afterward, the concept of concurrence was extended to two-qubit
mixed states by means of convex roof construction \cite{Wootters},
and then, to arbitrary but finite-dimensional bipartite as well as
multipartite systems for both pure and mixed states
\cite{Chen,Rungta}.

The continuous-variable systems can also be used for quantum
information processing and quantum computing \cite{Braunstein}. Most
analysis of entanglement in continuous-variable systems relies on
expressing the states of the system in terms of some discrete but
infinite basis. Then, the following problems arisen naturally: Can
the concept of concurrence be extended to infinite-dimensional case?
Is it also a ``well-defined'' entanglement measure? In the present
paper, we answer these questions affirmatively.

In \cite{ZW}, the \emph{partial Hermitian conjugate} (PHC) criterion
for  pure states in finite-dimensional systems was proposed and then
generalized in \cite{GY} to infinite-dimensional case. The (PHC)
criterion says that:
 A bipartite pure state is separable if and only if it is PHC invariant (see
below). The authors of \cite{GY,ZW} also pointed out that  one may
obtain an entanglement measure from (PHC) criterion since, for any
entangled pure state, the PHC of it is not equal to itself, and
thus the trace norm or the Hilbert-Schmidt norm of the difference
between them may be an entanglement measure of the given state.
Interestingly, as what we will show,  underlying the
Hilbert-Schmidt norm, the induced entanglement measure, \emph{PHC
measure}, coincides with concurrence. This result makes
contribution to the more profound understanding of the
concurrence.

In this paper, we consider the bipartite system consisting of two
parties A and B which are associated with the state spaces $H_A$ and
$H_B$, respectively, with $\dim H_A\otimes H_B\leq+\infty$. We
denote by $\rho_A$ and $\rho_B$ the reduced density operators of
$\rho$ with respect to the subsystems A and B, respectively, i.e.,
$\rho_A={\rm Tr}_B(\rho)$ and $\rho_B={\rm Tr}_A(\rho)$. A bipartite
state $\rho$ acting on $H=H_A\otimes H_B$ is called \emph{separable}
if it can be written as
\begin{eqnarray}
\rho=\sum\limits_ip_i\rho_i^A\otimes\rho_i^B,\quad
\sum\limits_ip_i=1, \ p_i\geq0
\end{eqnarray}
or it is a limit of the states of the above form under the trace
norm topology \cite{Werner}, where $\rho_i^A$ and $\rho_i^B$ are
pure states on the subsystems associated to the Hilbert spaces $H_A$
and $H_B$, respectively. A state that is not separable is said to be
\emph{entangled}. Particularly, if a state can be represented in the
form as in Eq.(1), it is called \emph{countably separable}
\cite{Holevo}. It is worth mentioning that, with increasing state
space dimension, quantifying entanglement becomes more and more
difficult to implement in practice.

The structure of this paper is as follows. In Sec.II we extend the
concept of the concurrence to infinite-dimensional bipartite systems
and show that it is a continuous function under the trace-class norm
topology. This result is new even for finite-dimensional case, and
enables  us to prove that the concurrence is also a well-defined
monotonic entanglement measure for infinite-dimensional case. Going
further, another entanglement measure which is closely related to
concurrence, \emph{tangle}, is investigated.  The PHC measure is
introduced and discussed in Sec.III. A brief conclusion is given in
the last section.

\section{Concurrence for infinite-dimensional bipartite states}

We start by reviewing some results from  finite-dimensional cases.
For the bipartite pure state $|\psi\rangle\in H_A\otimes H_B$ with
$\dim H_A\otimes H_B<+\infty$, the concurrence $C(|\psi\rangle)$ of
$|\psi\rangle$ is defined in \cite{Rungta} by
\begin{eqnarray}
C(|\psi\rangle)=\sqrt{2[1-{\rm Tr}(\rho_A^2)]},
\end{eqnarray}
where
$\rho_A={\rm Tr}_B(|\psi\rangle\langle\psi|)$.
Equivalently,
\begin{eqnarray*}
C(|\psi\rangle)
=\sqrt{\sum\limits_{i,j,k,l}|a_{ik}a_{jl}-a_{il}a_{jk}|^2}
\end{eqnarray*}
provided that
$|\psi\rangle=\sum\limits_{i,j}a_{ij}|i\rangle|j'\rangle$, where
$\{|i\rangle\}$ and $\{|j'\rangle\}$ are given orthonormal bases of
$H_A$ and $H_B$, respectively. The concurrence is extended to mixed
states by means of convex roof construction \cite{Rungta2},
\begin{eqnarray}
C(\rho)
=\min\limits_{\{p_i,|\psi_i\rangle\}}\{\sum\limits_i p_i
C(|\psi_i\rangle)\},
\end{eqnarray}
where the minimum is taken over all possible ensembles of $\rho$
(here, $\{p_i,|\psi_i\rangle\}$ is called an ensemble of $\rho$
whenever $\rho=\sum\limits_ip_i|\psi_i\rangle\langle\psi_i|$ with
$\{p_i\}$ a probability distribution and $\{|\psi_i\rangle\}$ a
family of pure states).

The tangle is another measure closely related to the concurrence.
The tangle $\tau(|\psi\rangle)$ for pure state $|\psi\rangle$ is
defined by $\tau(|\psi\rangle)=C^2(|\psi\rangle)$, and the tangle
for mixed state $\rho$ is defined by
\begin{eqnarray}
\tau(\rho)
=\min\limits_{\{p_i,|\psi_i\rangle\}}
\{\sum\limits_i p_i C^2(|\psi_i\rangle)\}
\end{eqnarray}
(Ref. \cite{Coffman}). Note that, although the tangle and the
concurrence are equivalent to each other as  entanglement measures
for pure states, they are different for mixed states. In fact, it
holds that $\tau(\rho)\geq C^2(\rho)$ and the equality holds in the
case of two-qubit states \cite{Osborne}. It is evident that $\rho$
is separable if and only if $C(\rho)=\tau(\rho)=0$.

With the same spirit in mind, we extend the concepts of concurrence
and tangle to infinite-dimensional bipartite systems. \\

\noindent {\bf Definition 1.} \ Let $|\psi\rangle\in H_A\otimes H_B$
with $\dim H_A\otimes H_B=+\infty$ be a pure state.
\begin{eqnarray}
C(|\psi\rangle):=\sqrt{2(1-{\rm Tr}(\rho_A^2))},
\end{eqnarray}
is called
the concurrence of $|\psi\rangle$, where $\rho_A={\rm Tr}_B(|\psi\rangle\langle\psi|)$.\\

Since the eigenvalues of $\rho_A$ coincide with that of $\rho_B={\rm
Tr}_A(|\psi\rangle\langle\psi|)$, with no loss of generality,
we always use the reduced density operators
with respect to the subsystem A. It is clear that
$C(|\psi\rangle)=0$ if and only if $|\psi\rangle$ is separable.

For a mixed state $\rho$, the concurrence of $\rho$ can be defined
by means of the generalized convex roof construction, namely,
\begin{eqnarray}
C(\rho):=\inf\limits_{\{p_i,|\psi_i\rangle\}}
\{\sum\limits_i p_i C(|\psi_i\rangle)\},
\end{eqnarray}
where the infimum is taken over all possible ensembles
$\{p_i,|\psi_i\rangle\}$ of $\rho$.

The following proposition provides two  computational formulas of
the concurrence for pure states.\\

\noindent{\bf Proposition 1} \  Let $|\psi\rangle\in H_A\otimes
H_B$ with $\dim H_A\otimes H_B=+\infty$ be a pure state.

(1)  If the Fourier expansion of $|\psi\rangle$ with respect to some
given product basis
 $\{|i\rangle|j'\rangle\}$
 of  $H_A\otimes H_B$ is
$|\psi\rangle=\sum\limits_{i,j}a_{ij}|i\rangle|j'\rangle$,
 then
\begin{eqnarray}
C(|\psi\rangle)
=\sqrt{\sum\limits_{i,j,k,l}|a_{ik}a_{jl}-a_{il}a_{jk}|^2}.
\end{eqnarray}

(2) If the Schmidt decomposition of  $|\psi\rangle$ is
$|\psi\rangle=\sum\limits_k\lambda_k|k\rangle|k'\rangle$, then
\begin{eqnarray}
C(|\psi\rangle)=\sqrt{2\sum\limits_{k\neq l}\lambda_k^2\lambda_l^2}.
\end{eqnarray}

\noindent{\sl Proof} \ (1) Consider the operator
$D=D_{|\psi\rangle}=(a_{ij}) : H_B\rightarrow H_A$ defined by
$D|j^\prime \rangle=\sum_i a_{ij}|i\rangle$. Since ${\rm
Tr}(DD^\dag)=\sum\limits_{i,j}|a_{ij}|^2=1$, $D$ is a
Hilbert-Schmidt operator. With $\rho=|\psi\rangle\langle\psi|$, it
is easily checked that $\rho_A=DD^\dag$.  As  ${\rm
Tr}((DD^\dag)^2)=\sum\limits_{i,j,k,l}a_{ik}\bar{a}_{il}a_{jl}\bar{a}_{jk}$,
 we have
\begin{eqnarray*}
&&1-{\rm Tr}(\rho_A^2)=(\sum\limits_{i,j}a_{ij}\bar{a}_{ij})^2-\sum\limits_{i,j,k,l}a_{ik}\bar{a}_{il}a_{jl}\bar{a}_{jk}\\
&=&\sum\limits_{i,j,k,l}
(a_{ik}\bar{a}_{ik}a_{jl}\bar{a}_{jl}-a_{ik}\bar{a}_{il}a_{jl}\bar{a}_{jk})\\
&=&\frac{1}{2}\sum\limits_{i,j,k,l}
(a_{ik}a_{jl}-a_{il}a_{jk})(\bar{a}_{ik}\bar{a}_{jl}-\bar{a}_{il}\bar{a}_{jk})\\
&=&\frac{1}{2}\sum\limits_{i,j,k,l}|a_{ik}a_{jl}-a_{il}a_{jk}|^2.
\end{eqnarray*}
Hence $C(|\psi\rangle)=\sqrt{2(1-{\rm Tr}(\rho_A^2))}
=\sqrt{\sum\limits_{i,j,k,l}|a_{ik}a_{jl}-a_{il}a_{jk}|^2}$, as
desired.

(2) can be checked similarly and we omit its proof here.\hfill$\qed$
\\

It is known that the concurrence is an entanglement measure for
finite dimensional systems since it meets the following
conditions:

(i) $E(\rho)=0$ if and only if $\rho$ is separable;

(ii) $E(\rho)=E(U_A\otimes U_B \rho U_A^\dag\otimes U_B^\dag)$ holds
for any local unitary operators $U_A$ and $U_B$ on the subsystems
$H_A$ and $H_B$, respectively;

(iii) $E$ is LOCC monotonic, i.e., $E(\Lambda(\rho))\leq E(\rho)$
holds for any local operation and classical communication (LOCC)
$\Lambda$ \cite{Fan}.

The conditions (i)-(iii) above are necessary for any entanglement
measure $E$ \cite{Vedral}. Generally, an entanglement measure $E$
also satisfies

(iv) $E(\sum\limits_ip_i\rho_i)\leq\sum\limits_ip_iE(\rho_i)$ for
mixed state $\rho=\sum\limits_ip_i\rho_i$, where $p_i\geq0$,
$\sum\limits_ip_i=1$ (see in \cite{Vedral2}).

If (iii)-(iv) are satisfied by an entanglement measure $E$, then it
is called an entanglement monotone \cite{Vidal}.

In what follows we show that the concurrence $C$ defined in
Eq.(4)-(5) for infinite-dimensional systems is also an
entanglement monotone, i.e., (i)-(iv) are satisfied by $C$ for
infinite-dimensional case as well.

Checking $C$ meets Condition (ii) is straightforward.

The condition (i) is obviously satisfied by the concurrence for
finite-dimensional case.  This is because that,
 every separable state $\rho$ in a finite-dimensional bipartite system is countably separable,
 that is, there  exists an
ensemble $\{p_i,|\psi_i\rangle\}$ of $\rho$ such that
$|\psi_i\rangle$s are separable pure states  and thus we get
immediately that $0\leq C(\rho)\leq \sum_i p_iC(|\psi_i\rangle)=0$
as  $C(|\psi_i\rangle)=0$. However, the fact that $\rho$ is
separable implies $C(\rho)=0$  is not obvious anymore for
infinite-dimensional case since   there do exist some separable
states in infinite-dimensional systems that are not countably
separable \cite{Holevo}. For such  separable states that are not
\emph{countably separable}, there doesn't exist any ensemble
$\{p_i,|\psi_i\rangle\}$ of $\rho$ such that $|\psi_i\rangle$s are
separable and one can not get $C(\rho)=0$ directly. It is clear
that, if $C$ is continuous, then $C(\rho)=0$ whenever $\rho$ is
separable because it is a limit of countably separable states.

The continuity of the concurrence $C$ is established in Proposition
2,
which is not obvious even for finite-dimensional systems.\\

\noindent{\bf Proposition 2} \ The concurrence is continuous for
both finite- and infinite-dimensional systems, i.e.,
\begin{eqnarray}
\lim_{n\rightarrow\infty}C(\rho_n) =C(\rho)\quad {\rm
whenever}\quad \lim_{n\rightarrow\infty}\rho_n=\rho
\end{eqnarray}
in the trace-norm topology.\\

\noindent{\sl Proof} \ To prove the continuity of $C$, let us extend
the concurrence of  states to that of self-adjoint trace-class
operators.

Let $A$ be a self-adjoint trace-class operators acting on
$H_A\otimes H_B$. We define the concurrence of $A$ by
$$ C(A)={\rm Tr}(|A|)C(\frac{|A|}{{\rm Tr}(|A|)}),$$
where $|A|=(A^\dag A)^{\frac{1}{2}}$. It is clear that
$$C(A)=\inf\limits_{\{\lambda_i, |\psi_i\rangle\}}
\sum_i\lambda_iC(|\psi_i\rangle),$$ where the infimum is taken over
all $\{\lambda_i, |\psi_i\rangle\}$ with $\lambda_i\geq 0$,
$\sum_i\lambda_i={\rm Tr}(|A|)$ and
$|A|=\sum_i\lambda_i|\psi_i\rangle\langle\psi_i|$. It is an
immediate consequence of the definition that, if $0\leq |A|\leq
|B|$, then $C(A)\leq C(B)$.

 Assume that $\rho_n,\rho\in {\mathcal S}(H_A\otimes H_B)$ and  $\lim_{n\rightarrow\infty}\rho_n=\rho$.
 Let $\vartheta_n=\rho-\rho_n$ and let
\begin{eqnarray*}\vartheta_n
=\sum_{k(n)}\lambda_{k(n)}|\eta_{k(n)}\rangle\langle\eta_{k(n)}|
\end{eqnarray*}
be its spectral decomposition.

We claim that
\begin{eqnarray}
C(\rho)=C(\rho_n+\vartheta_n)\leq C(\rho_n)+C(\vartheta_n).
\end{eqnarray}
For any  $\varepsilon>0$, there exist ensembles $\{p_{k(n)}$,
$|\psi_{k(n)}\rangle\}$ and $\{q_{l(n)}$, $|\phi_{l(n)}\rangle\}$ of
$\rho_n$ and ${|\vartheta_n|}$, respectively,  and
$0<\epsilon_1,\epsilon_2< \varepsilon$, such that
\begin{eqnarray*}C(\rho_n)=\sum_{k(n)}p_{k(n)}C(|\psi_{k(n)}\rangle)-\frac{\epsilon_1}{2}
\end{eqnarray*}
and
\begin{eqnarray*}
C({|\vartheta_n|})=\sum_{l(n)}q_{l(n)}C(|\phi_{l(n)}\rangle)-\frac{\epsilon_2}{2}.
\end{eqnarray*}
We compute
\begin{eqnarray*}
&&C(\rho_n+\vartheta_n)\leq C(\rho_n+|\vartheta_n|)\\
&\leq&\sum_{k(n)}p_{k(n)}C(|\psi_{k(n)}\rangle)
+\sum_{l(n)}q_{l(n)}C(|\phi_{l(n)}\rangle)\\
&=&C(\rho_n)+C(\vartheta_n)+\frac{\epsilon_1+\epsilon_2}{2}.
\end{eqnarray*}
Since $\varepsilon$ is arbitrarily given, the claim is proved.

Similarly, using $C(\rho_n)=C(\rho-\vartheta_n)\leq
C(\rho+|\vartheta_n|)$, we  obtain
\begin{eqnarray*}
C(\rho_n)\leq C(\rho)+C(|\vartheta|_n),
\end{eqnarray*}
which, together with Eq.(10), implies that
\begin{eqnarray*}
|C(\rho_n)-C(\rho)|\leq C(|\vartheta|_n).
\end{eqnarray*}
Observing that $C(\vartheta_n)\rightarrow0$ $(n\rightarrow\infty)$
since $C(\vartheta_n)\leq\sum_{k(n)}\sqrt{2}|\lambda_{k(n)}|$ and
${\rm Tr}(|\vartheta_n|) =\sum_{k(n)}|\lambda_{k(n)}|\rightarrow0$,
we get $\lim_{n\rightarrow\infty}C(\rho_n)=C(\rho)$, as desired.
\hfill$\qed$\\

We now begin to check that $C$ satisfies properties (iii)-(iv). For
finite-dimensional case, Vidal \cite{Vidal} proposed a nice recipe
for determining entanglement monotones by proving that the convex
roof extension of a pure sate measure $E$ satisfying the two
conditions below is an entanglement monotone  (Ref.  \cite[Theorem
2]{Vidal}):\\

(a) For a pure state $|\psi\rangle$, $\rho_A={\rm
Tr}_B(|\psi\rangle\langle\psi|)$, define a function $f$ by
$f(\rho_A)=E(|\psi\rangle)$, then
$$f(U\rho_AU^\dag)=f(\rho_A);$$ and

(b) $f$ is concave, namely,
$$f(\lambda\rho_1+(1-\lambda)\rho_2)\geq\lambda f(\rho_1)+(1-\lambda)f(\rho_2)$$
for any density matrices $\rho_1$, $\rho_2$, and any
$0\leq\lambda\leq1$.\\

For infinite-dimensional bipartite systems,
every LOCC  admits a form
of
\begin{eqnarray}
\Lambda(\rho) =\sum\limits_{i=1}^N(A_i\otimes B_i)\rho (A_i^\dag
\otimes B_i^\dag)
\end{eqnarray}
 with $\sum\limits_{i=1}^NA_i^\dag
A_i\otimes B_i^\dag B_i\leq I_A\otimes I_B$,  where $N$ may be
$+\infty$ and   the series converges in the strong operator topology
\cite{HJ}. \if false It is proved that, any LOCC $\Lambda$ of
bipartite pure state using two-way communication can be simulated by
a one-way communication protocol \cite{Lo}. Namely, $\Lambda$ can be
simulated by
$\Lambda'(|\psi\rangle\langle\psi|)=\sum\limits_{i}(I_A\otimes
\Lambda_i^B)(A_i\otimes I_B|\psi\rangle\langle\psi| A_i^\dag\otimes
I_B)$, where $\Lambda_i^B$s are quantum operations on the second
system [namely, $\Lambda_i^B$ is a completely positive trace
preserving linear map, it admits a form of
$\Lambda_i^B(\cdot)=\sum\limits_k M_{k(i)}(\cdot)M_{k(i)}^\dag$ with
$\sum\limits_{k(i)}M_{k(i)}^\dag M_{k(i)}=I_B$, where the series
converges in the strong operator topology (see \cite{HJ})],
$\sum\limits_i A_i^\dag A_i\leq I_A$ and $\Lambda_i^B$s are quantum
operations conditional on the result $i$, where the series converges
in the strong operator topology \cite{HJ}. \fi  Let
$\mathcal{S}(H_A\otimes H_B)$ be the set of all quantum states
acting on $H_A\otimes H_B$. According to the entanglement monotone
scenario discussed in \cite{Vidal}, in order to prove that a
function  $E: {\mathcal S}(H_A\otimes H_B)\rightarrow{\mathbb
{R}}^+$
 satisfying (i)-(ii) is LOCC monotonic, we only need to consider
the sequence of LOCC  $\{\Lambda_{B,k}\}$ or $\{\Lambda_{A,l}\}$, of
the form
\begin{eqnarray}
\Lambda_{B,k}(\rho)=\sum_{i(k)}(I_A\otimes B_{i(k)})\rho(I_A\otimes
B_{i(k)}^\dag)
\end{eqnarray}
or
\begin{eqnarray}
\Lambda_{A,l}(\rho)=\sum_{j(l)}(A_{j(l)}\otimes
I_B)\rho(A_{j(l)}^\dag\otimes I_B),
\end{eqnarray}
where $\sum_{i(k)}B_{i(k)}^\dag B_{i(k)}\leq I_B$ and
$\sum_{j(l)}A_{j(l)}^\dag A_{j(l)}\leq I_A$ (here, the series
converges in the strong operator topology) with $\sum\limits_k{\rm
Tr}(\Lambda_{B,k}(\rho))=\sum\limits_l{\rm
Tr}(\Lambda_{A,l}(\rho))=1$, $B_{i(k)}$s (resp. $A_{j(l)}$s) are
operators from $H_B$ (resp. $H_A$) into $H_{B'}$ (resp. $H_{A'}$)
for some Hilbert space $H_{B'}$ (resp. $H_{A'}$), and where $k$
(resp. $l$) labels different outcomes if at some stage of local
manipulations part B (resp. A) performs a measurement. With no loss
of generality, hereafter we consider the LOCC $\{\Lambda_{B,k}\}$ as
in Eq.(12). Applying $\Lambda_{B,k}$ to $\rho$, the state becomes
$$\rho_k'=\frac{\Lambda_{B,k}(\rho)}{p_k}$$
with probability $p_k={\rm Tr}(\Lambda_{B,k}(\rho))$. Therefore, the
final state is $\rho'=\sum_kp_k\rho_k'$. By \cite{Vidal}, if
$E(\rho')\leq E(\rho)$ holds for $\Lambda_{B,k}$, then the condition
(iii) holds for $E$. We show below that, for infinite-dimensional
case, if $E$ is continuous on quantum states under the trace norm
topology, then (a)-(b) are sufficient conditions for $E$
to be an entanglement monotone as well.\\

\noindent{\bf Proposition 3} \  Let $E$ be an entanglement measure
for pure states in infinite-dimensional systems and define
$E(\rho):$=$\inf\limits_{\{p_i,|\psi_i\rangle\}}$ $\{\sum\limits_i
p_i E(|\psi_i\rangle)\}$ for mixed state $\rho$. Let
$f(\rho_A)=E(|\psi\rangle\langle\psi|)$, $\rho_A={\rm
Tr}_B(|\psi\rangle\langle\psi|)$. Assume that $E$ is continuous and
$f$ satisfying (a)-(b).  Then $E$ is an entanglement monotone, i.e.,
$E$ satisfying (iii)-(iv).\\

\noindent{\sl Proof} \ We assume that $f$ satisfy conditions (a) and
(b), namely (a) $f(U\rho_AU^\dag)=f(\rho_A)$ for any unitary
operators on $H_A$ and (b) $f$ is concave.

By (a), we know that $E(\rho)$ is invariant under local unitary
operations, i.e., $E(U_A\otimes U_B\rho U_A^\dag\otimes
U_B^\dag)=E(\rho)$ for any unitary operators $U_A$ and $U_B$ acting
on $H_A$ and $H_B$ respectively (notice that condition (iii) implies
that $E(\rho)$ is invariant under local unitary operations).

In what follows, we show that (iii) holds  for $E$ and LOCC
$\{\Lambda_{B,k}\}$, from which, according to the entanglement
monotone scenario proposed in \cite{Vidal}, we can thus obtain that
(iii) holds for $E$ and any LOCC $\Lambda$.

We assume first that $\rho$ is a pure state,
$\rho=|\psi\rangle\langle\psi|$. If part B performs $\Lambda_{B,k}$
on subsystem B as in Eq.(11), then the state becomes
$\rho_k'=\frac{\Lambda_{B,k}(\rho)}{p_k}$ with probability $p_k={\rm
Tr}(\Lambda_{B,k}(\rho))$. Writing $\rho_{A,k}'={\rm Tr}_B(\rho_k)$,
we obtain $\rho_A=\sum_kp_k\rho_{A,k}'$. For any ensemble
$\{r_{kl},|\psi_{kl}\rangle\}$ of $\rho_k'$, we have
$$E(\rho_k')\leq\sum_lr_{kl}E(|\psi_{kl}\rangle).$$
It yields
\begin{eqnarray*}
&&E(\rho)=f(\rho_A)=f(\sum_kp_k\rho_{A,k}')\\
&=&f(\sum_{k,l}p_kr_{kl}\rho_{A,kl}')
\geq\sum_{k,l}p_kr_{kl}f(\rho_{A,kl}')\\
&=&\sum_{k,l}p_kr_{kl}E(|\psi_{kl}\rangle) \geq\sum_kp_kE(\rho_k'),
\end{eqnarray*}
where $\rho_{A,kl}'={\rm Tr}_B(|\psi_{kl}\rangle\langle\psi_{kl}|)$,
the first inequality holds since $f$ is concave and continuous.
Therefore, (iii) is satisfied by $E$ if $\rho$ is pure.

Assume that $\rho$ is mixed. Performing  $\Lambda_{B,k}$ on $\rho$
and denote $\rho_k'=\frac{\Lambda_{B,k}(\rho)}{p_k}$ with
probability $p_k={\rm Tr}(\Lambda_{B,k}(\rho))$. Observe that, for
any  $\varepsilon>0$, there exists an ensemble
$\{t_j,|\eta_j\rangle\}$ of $\rho$, and $0<\epsilon_1<\varepsilon$
such that
\begin{eqnarray*}
E(\rho)=\sum_jt_jE(|\eta_j\rangle)-\frac{\epsilon_1}{2}.
\end{eqnarray*}
For each $j$, let
\begin{eqnarray*}
\rho_{jk}'=\frac{1}{t_{jk}}\Lambda_{B,k}(|\eta_j\rangle\langle\eta_j|),
\end{eqnarray*}
where $t_{jk}={\rm
Tr}(\Lambda_{B,k}(|\eta_j\rangle\langle\eta_j|))$. Then
\begin{eqnarray*}
\rho_k'=\frac{1}{p_k}\sum_jt_jt_{jk}\rho_{jk}'
\end{eqnarray*}
and
\begin{eqnarray*}
E(|\eta_j\rangle)\geq\sum_kt_{jk}E(\rho_{jk}')
\end{eqnarray*}
by what proved for pure states above.  For each pair $(j,k)$,
suppose that $\{t_{jkl},|\psi_{jkl}\rangle\}$ is an ensemble of
$\rho_{jk}'$ such that
\begin{eqnarray*}
E(\rho_{jk}')=\sum_lt_{jkl}E(|\psi_{jkl}\rangle)-\frac{\epsilon_{jk}}{2},\quad
0<\epsilon_{jk}<\frac{\varepsilon}{2^k}.
\end{eqnarray*}
We achieve that
\begin{eqnarray*}
E(\rho)&=&\sum_jt_jE(|\eta_j\rangle)-\frac{\epsilon_1}{2}\\
&\geq&
\sum_{j,k}t_jt_{jk}E(\rho_{jk}')-\frac{\epsilon_1}{2}\\
&=&\sum_{j,k,l}t_jt_{jk}t_{jkl}
E(|\psi_{jkl}\rangle)-\epsilon'\\
&\geq&\sum_kp_kE(\rho_k')-\epsilon'
\end{eqnarray*}
for some $\epsilon'<\varepsilon$. Since $\varepsilon$ is arbitrarily
given, we see that (iii) is satisfied for mixed states as well.

Now we show that (iv) is valid. Let $\rho=\sum_kp_k\rho_k$. For any
given $\varepsilon>0$, there exists an ensemble of $\rho_k$,
$\{q_{kl},|\phi_{kl}\rangle\}$, and $0<\epsilon<\varepsilon$ such
that
\begin{eqnarray*}
E(\rho_k)\geq\sum_lq_{kl}E(|\phi_{kl}\rangle)-\frac{\epsilon}{2^k}.
\end{eqnarray*}
As $\{p_kq_{kl},|\phi_{kl}\rangle \}_{k,l}$ is an ensemble of
$\rho$, this entails that
\begin{eqnarray*}E(\rho)\leq\sum_kp_k\sum_lq_{kl}E(|\phi_{kl}\rangle)\leq\sum_kp_kE(\rho_k)+{\varepsilon},
\end{eqnarray*}
 from which we see that $E(\rho)\leq\sum_kp_kE(\rho_k) $,
finishing the proof. \hfill$\qed$ \\

Based on Proposition 3, we show below that the concurrence for infinite-dimensional
systems defined in Eqs.(4)-(5) satisfies conditions (i)-(iv) and thus it
is a well-defined entanglement measure (monotone).\\

\noindent{\bf Theorem 1} \  The concurrence defined in Eqs.(4)-(5) is
an entanglement monotone.\\

\noindent{\sl Proof} \ By Proposition 2 and Proposition 3, we only
need to verify that the function $f$ defined by
$f(\rho_A)=C(|\psi\rangle)$ with $\rho_A={\rm
Tr}_B(|\psi\rangle\langle\psi|)$ satisfies (a) and (b). Note that,
for any $\rho\in{\mathcal S}(H_A)$, we have $f(\rho)=\sqrt{2(1-{\rm
Tr}(\rho^2))}$. Thus (a) is obvious. We check that $f$ is concave.
For any given states $\rho_1$ and $\rho_2$ on $H_A$, let
\begin{eqnarray*}
\rho=\lambda\rho_1+(1-\lambda)\rho_2,\quad 0\leq\lambda\leq1.
\end{eqnarray*}
Then
\begin{eqnarray*}
f(\lambda\rho_1+(1-\lambda)\rho_2)\geq\lambda f(\rho_1)+(1-\lambda)f(\rho_2)
\end{eqnarray*}
if and only if \if false
\begin{eqnarray}
{\rm Tr}(\rho_1\rho_2)
+\sqrt{(1-{\rm Tr}(\rho_1^2))(1-{\rm Tr}(\rho_1^2))}\leq1.
\end{eqnarray}
Using the fact
$(\rho_1-\rho_2)(\rho_1-\rho_2)^\dag=(\rho_1-\rho_2)^2=\rho_1^2+\rho_2^2-\rho_1\rho_2-\rho_2\rho_1\geq0$,
one has\fi
\begin{eqnarray*}
{\rm Tr}(\rho_1^2)+{\rm Tr}(\rho_2^2)\geq2{\rm Tr}(\rho_1\rho_2).
\end{eqnarray*}
But the last inequality is always valid. Thus, $f$ is concave.
 \hfill$\qed$\\

With the same spirit as that for   finite-dimensional case, we
define the tangle of a pure state in the case of
infinite-dimensional systems by
\begin{eqnarray}
\tau(|\psi\rangle)=C^2(|\psi\rangle).
\end{eqnarray}
If $|\psi\rangle=\sum\limits_k\lambda_k|k\rangle|k'\rangle$ is the
Schmidt decomposition of $|\psi\rangle$ \cite{GY}, then
\begin{eqnarray}
\tau(|\psi\rangle)=2(1-{\rm Tr}(\rho_A^2))=2\sum\limits_{k\neq
l}\lambda_k^2\lambda_l^2.
\end{eqnarray}
The tangle of a mixed state $\rho$, $\tau(\rho)$ can be naturally
defined by
\begin{eqnarray}
\tau(\rho):=\inf\limits_{\{p_i,|\psi_i\rangle\}}
\{\sum\limits_i p_i C^2(|\psi_i\rangle)\},
\end{eqnarray}
where the infimum is taken over all possible ensembles
$\{p_i,|\psi_i\rangle\}$ of $\rho$. By Proposition 2 and Theorem 1,
$\tau$ is continuous and   satisfies the conditions (i)-(iv) as
well. Therefore, $\tau$ is an good entanglement measure, too.

For mixed state $\rho$, $C(\rho)\neq \sqrt{2[1-{\rm Tr}(\rho_A^2)]}$
in general. For the finite-dimensional case, it is showed in
\cite{Zhang,Mintert} that $C^2(\rho)\leq2[1-{\rm Tr}(\rho_A^2)]$. In
fact, we have the following result.\\

\noindent{\bf Proposition 4} \ Let $\rho\in{\mathcal S}(H_A\otimes
H_B)$ with $\dim H_A\otimes H_B\leq\infty$. Then
\begin{eqnarray}
C^2(\rho)\leq\tau(\rho)\leq2[1-{\rm Tr}(\rho_A^2)].
\end{eqnarray}

\noindent{\sl Proof} \ For any $\epsilon>0$, there exists
$\{p_i,|\psi_i\rangle\}$ such that
$\rho=\sum\limits_ip_i|\psi_i\rangle\langle\psi_i|$ and
\begin{eqnarray*}
\tau(\rho)\geq\sum\limits_ip_iC^2(|\psi_i\rangle)-\epsilon.
\end{eqnarray*}
 Then we have
\begin{eqnarray*}
&&C^2(\rho)\leq(\sum\limits_ip_iC(|\psi_i\rangle))^2\\
&=&(\sum\limits_i\sqrt{p_i}\sqrt{p_i}C(|\psi_i\rangle))^2\\
&\leq&(\sum\limits_ip_i)(\sum\limits_ip_iC^2(|\psi_i\rangle))\\
&\leq&\tau(\rho)+\epsilon,
\end{eqnarray*}
which establishes the inequality $C^2(\rho)\leq\tau(\rho)$ since
$\varepsilon>0$ is arbitrary.

 Let
$\rho_{i,A}={\rm Tr}_B(|\psi_i\rangle\langle\psi_i|)$. One has
\begin{eqnarray*}\tau(\rho)&\leq &\sum\limits_ip_iC^2(|\psi_i\rangle)\\
&=&\sum\limits_ip_i[2(1-{\rm Tr}(\rho_{i,A}^2))]\\
&=&2(1-\sum\limits_ip_i{\rm Tr}(\rho_{i,A}^2))\\
&\leq&2(1-{\rm Tr}(\rho_A^2))
\end{eqnarray*}
due to the convex property of ${\rm Tr}(\rho_A^2)$ \cite{Zhang}.  \hfill$\qed$\\

\section{PHC measure: viewing concurrence from another perspective
}

In this section, we will establish another entanglement measure--PHC
measure which is based on the PHC criterion \cite{ZW,GY} and show
that it coincides with concurrence. Therefore, it provides us an
alternative perspective of understanding  the concurrence.

It is known that entanglement measures may be induced from some
entanglement criteria. For example, negativity and convex-roof
extended negativity are two kinds of entanglement measures induced
from the elegant PPT criterion \cite{Vidal3,Lee}. In \cite{GY,ZW}, a
necessary and sufficient condition of separability for pure states
is proposed. We review some notations firstly.
\\

\noindent{\bf Definition 2}(\cite{ZW,GY}). \  Let
$\rho=|\psi\rangle\langle\psi|$ be a pure state acting on
$H_A\otimes H_B$ with $\dim H_A\otimes H_B\leq+\infty$ and
\begin{eqnarray*}
|\psi\rangle=\sum\limits_k\lambda_k|k\rangle|k^{\prime}\rangle
\end{eqnarray*}
be the
Schmidt decomposition of $|\psi\rangle$. Then
\begin{eqnarray*}
\rho=\sum\limits_{k,l}\lambda_k\lambda_l|k\rangle|k^{\prime}\rangle\langle
l|\langle l^{\prime}|
\end{eqnarray*}
and the partial Hermitian conjugate of $\rho$ is
defined by
\begin{eqnarray}\rho^{\rm PHC}=\sum\limits_{k,l}\lambda_k\lambda_l|k\rangle|l^{\prime}\rangle
\langle l|\langle k^{\prime}|.
\end{eqnarray}

It is showed in \cite{ZW,GY} that a pure state
$\rho=|\psi\rangle\langle\psi|$, $|\psi\rangle\in H_A\otimes H_B$,
is separable if and only if $\rho^{\rm PHC}=\rho$.  Consequently,
$\rho^{\rm PHC}\neq\rho$ implies that $\rho$ is entangled, and also
that $\|\rho_\psi-\rho_\psi^{\rm PHC}\|_2>0$ (here, $\|\cdot\|_2$
denotes the Hilbert-Schmidt norm, i.e., $\|A\|_2=[{\rm Tr}(A^\dag
A)]^{\frac{1}{2}}$). In what follows, we will show that
the PHC criterion does provide us with an entanglement measure.\\

\noindent{\bf Definition 3} \  Let
$|\psi\rangle\in H_A\otimes H_B$ with $\dim H_A\otimes H_B\leq+\infty$ be a pure state.
The PHC measure of a pure state
$|\psi\rangle$ is defined by
\begin{eqnarray}
E_{\rm
PHC}(|\psi\rangle):=\|\rho_\psi-\rho_\psi^{\rm PHC}\|_2.
\end{eqnarray}
For mixed state $\rho$,
\begin{eqnarray}
E_{\rm PHC}(\rho)
:=\inf\limits_{\{p_i,|\psi_i\rangle\}}
\{\sum\limits_i p_i E_{\rm PHC}(|\psi_i\rangle)\},
\end{eqnarray}
where the infimum is taken over all possible ensembles of $\rho$.

The main result of this section is the following.\\

\noindent{\bf Theorem 2} \  The PHC entanglement measure coincides
with the concurrence, i.e., $E_{\rm PHC}(\rho)=C(\rho)$ for any
state $\rho$ acting on $H_A\otimes H_B$ with
$\dim H_A\otimes H_B\leq+\infty$.\\

\noindent{\sl Proof} \ By the generalized convex roof construction, we only need to show
\begin{eqnarray}
E_{\rm PHC}(|\psi\rangle)=C(|\psi\rangle)
\end{eqnarray}
 for all pure
states $|\psi\rangle$.

Let $|\psi\rangle=\sum\limits_k\lambda_k|k\rangle |k^{\prime}\rangle$ be the
Schmidt decomposition of $|\psi\rangle$. Then
$\rho^{\rm PHC}=\sum\limits_{k,l}\lambda_k\lambda_l|k\rangle|l^{\prime}\rangle
\langle l|\langle k^{\prime}|$.
Therefore
\begin{eqnarray*}
&&\rho_\psi-\rho_\psi^{\rm
PHC}\\
&=&\sum\limits_{k,l}\lambda_k\lambda_l|k\rangle|k^{\prime}\rangle\langle
l|\langle l^{\prime}|-\sum\limits_{i,j}\lambda_i\lambda_j|i\rangle |j^{\prime}\rangle
\langle j|\langle i^{\prime}|\\
&=&\sum\limits_{k,l}\lambda_k\lambda_l|k\rangle\langle l|
\otimes(|k^\prime\rangle\langle l^\prime|-|l^\prime\rangle\langle
k^\prime|).
\end{eqnarray*}
 Now,
\begin{eqnarray*}
&&(\rho_\psi-\rho_\psi^{\rm
PHC})(\rho_\psi-\rho_\psi^{\rm
PHC})^\dagger\\
&=&\sum\limits_{k,l,i}\lambda_k\lambda_l^2\lambda_i|k\rangle\langle
i|\otimes(|k^\prime\rangle\langle i^\prime|+\langle
k^\prime|i^\prime\rangle|l^\prime\rangle\langle l^\prime|\\
 &&-\langle
l^\prime|i^\prime\rangle|k^\prime\rangle\langle l^\prime| -\langle
k^\prime|i^\prime\rangle|i^\prime\rangle\langle
i^\prime|)
\end{eqnarray*}
implies that
\begin{eqnarray*}
{\rm Tr}((\rho_\psi-\rho_\psi^{\rm
PHC})(\rho_\psi-\rho_\psi^{\rm PHC})^\dagger)=2\sum\limits_{k\neq
l}\lambda_k^2\lambda_l^2.
\end{eqnarray*}
 Therefore
\begin{eqnarray*}
E_{\rm
PHC}(|\psi\rangle)=\|\rho_\psi-\rho_\psi^{\rm
PHC}\|_2=\sqrt{2\sum\limits_{k\neq
l}\lambda_k^2\lambda_l^2}=C(|\psi\rangle)
\end{eqnarray*}
by Proposition 1, completing the proof. \hfill$\qed$\\

Thus, the PHC measure can also be regarded as a ``well-defined''
entanglement measure. Although the PHC measure is the same to the
concurrence, it shines some new light on the nature of the
concurrence.

\section{Conclusion}

Summarizing, the concepts of the concurrence and the tangle for
infinite-dimensional bipartite quantum systems are introduced. These
two functions are continuous under the trace norm topology.  This
enables us to prove that the concurrence as well as the tangle are
still well-defined monotonic entanglement measures. The relationship
between them are discussed and an upper bound is proposed:
$C(\rho)\leq\sqrt{\tau(\rho)}\leq\sqrt{2[1-{\rm Tr}(\rho_A^2)]}$,
where the equalities hold whenever $\rho$ is a pure state. Based on
the partial Hermitian conjugate criterion, the PHC measure is
introduced.  Moreover, this measure  coincides with the concurrence
and thus a well defined entanglement measure, which answers a
question suggested in \cite{ZW,GY}.\\

{\bf Acknowledgements.}\  This work is partially supported by
Natural Science Foundation of China (11171249,11101250) and Research
start-up fund for the Doctor of Shanxi Datong University.

\end{document}